\documentclass{iaus}
\usepackage{graphics}

  \checkfont{eurm10}
  \iffontfound
    \IfFileExists{upmath.sty}
      {\typeout{^^JFound AMS Euler Roman fonts on the system,
                   using the 'upmath' package.^^J}%
       \usepackage{upmath}}
      {\typeout{^^JFound AMS Euler Roman fonts on the system, but you
                   dont seem to have the}%
       \typeout{'upmath' package installed. iaus.cls can take advantage
                 of these fonts,^^Jif you use 'upmath' package.^^J}%
      }
  \else
  \fi

  \checkfont{msam10}
  \iffontfound
    \IfFileExists{amssymb.sty}
      {\typeout{^^JFound AMS Symbol fonts on the system, using the
                'amssymb' package.^^J}%
       \usepackage{amssymb}%

      }{}
  \fi

  \IfFileExists{amsbsy.sty}
    {\typeout{^^JFound the 'amsbsy' package on the system, using it.^^J}%
     \usepackage{amsbsy}}
    {}

\newsavebox{\astrutbox}
\sbox{\astrutbox}{\rule[-5pt]{0pt}{20pt}}

\title[Transition nuclei as evolved Composite nuclei]{Transition (LINER/HII) nuclei as evolved Composite (Seyfert 2/Starburst) nuclei}

\author[T. Storchi-Bergmann {\it et al.\/}]
{Thaisa Storchi-Bergmann$^1$,
C. H. Brandt$^1$, R. Cid Fernandes$^2$,\break 
H. R. Schmitt$^3$
\and R. Gonz\'alez Delgado$^4$}

\affiliation{$^1$Instituto de F\'\i sica, UFRGS, Porto Alegre, RS, Brazil
e-mail: thaisa@if.ufrgs.br\\[\affilskip]
$^2$Departamento de F\'\i sica, CFM -- UFSC, Florian\'opolis, SC, Brazil
e-mail: cid@astro.fs.ufsc.br\\[\affilskip]
$^3$National Radio Astronomy Observatory, Charlottesville, VA22903, USA
\\[\affilskip]
$^4$Instituto de Astrof\'\i sica de Andaluc\'\i a (CSIC), Granada 18080, Spain
e-mail: rosa@iaa.es}

\pubyear{2004}
\volume{222}
\pagerange{1--8}
\date{?? and in revised form ??}
\jname{The Interplay among Black Holes, Stars and ISM \\ in Galactic Nuclei}
\editors{T. Storchi-Bergmann, L. C. Ho \& H. R. Schmitt, eds.}
\begin{document}

\maketitle

\begin{abstract}
We compare the circumnuclear stellar population and environmental properies 
of Seyfert and  Composite (Seyfert + Starburst) nuclei with those 
of LINERs and LINER/HII transition galaxies (TOs), and discuss evidence
for evolution from Seyfert/Composite to LINER/TO nuclei.
\end{abstract}

\firstsection 
\section{Introduction}

Many contributions to these Proceedings discuss
signatures of the feeeding process of
AGN. Our group, in particular,
has been looking for peculiarities in the circumnuclear stellar
population in these galaxies.
We have concluded that about 40\% of nearby Seyferts
seem to have circumnuclear starbursts
(\cite{Cid01}),  which is significantly more than
what is found in a sample of non-active galaxies
with similar Hubble types, arguing for a relation
between the presence of a starburst and an active
nucleus. We have called the Seyfert galaxies with
luminous circumnuclear starbursts of Composites.

In \cite{SB01} we proposed an evolutionary
scenario for the activity in Seyferts in which
interactions with companion galaxies could be
the triggers of both nuclear activity and
circumnuclear starbursts. In this scenario the
age of the last generation of stars could be used
to date the onset of nuclear activity, if the
active nucleus and circumnuclear starburst
are triggered together. If, in addition,
they evolve together, we should find weak
active nuclei as evolved Seyfert 2 and old
starbursts around weak active nuclei as the
evolved Composite nuclei. These evolved nuclei
could be LINERs and TOs (\cite{Ho97}).
In this contribution we discuss evidence pointing in this direction.

\section{Stellar population {\it vs.} galaxy morphology and environment}

We first summarize the results found by \cite{SB01}, which motivated the proposed
evolutionary scenario:
(1) there is a larger incidence of Composites in late-type  galaxies; 
(2) the inner morphology of Composites is also ``late-type" (\cite{M98});
(3) the age of the starbursts in the Composites is in the range
$10^6-10^8$yrs;
(4) there is a larger incidence of close companions among Composites.
It is interesting to point out that these conclusions are similar to 
those reached by \cite{MT02} for a sample
of 48 CfA Seyfert galaxies, in particular the larger incidence of younger
stellar population in later type galaxies and in galaxies with close 
companions. In the evolutionary scenario the Composite nuclei would be the
youngest Seyferts, in which signatures of the interaction could still
be clearly visible, as well as there would be plenty of gas and dust,
which would lead to the late-type morphology.
 
Recently, we extended the stellar population study to
a sample  of 51 LINERs and TO's (\cite{Cid04}; \cite{R04}).
Investigating the relation  between  
the stellar population and environmental properties, we find that: 
(1) the stellar population of most TO's is characterized by a
significant contribution from intermediate age stars;
(2) there is a larger incidence of TOs in late-type galaxies; 
(3) the age of the starbursts is mostly around 10$^9$yrs;
(4) there is no excess of companions among TOs (Schmitt 2001).

\begin{figure}
\centering
\resizebox{12cm}{!}{\includegraphics{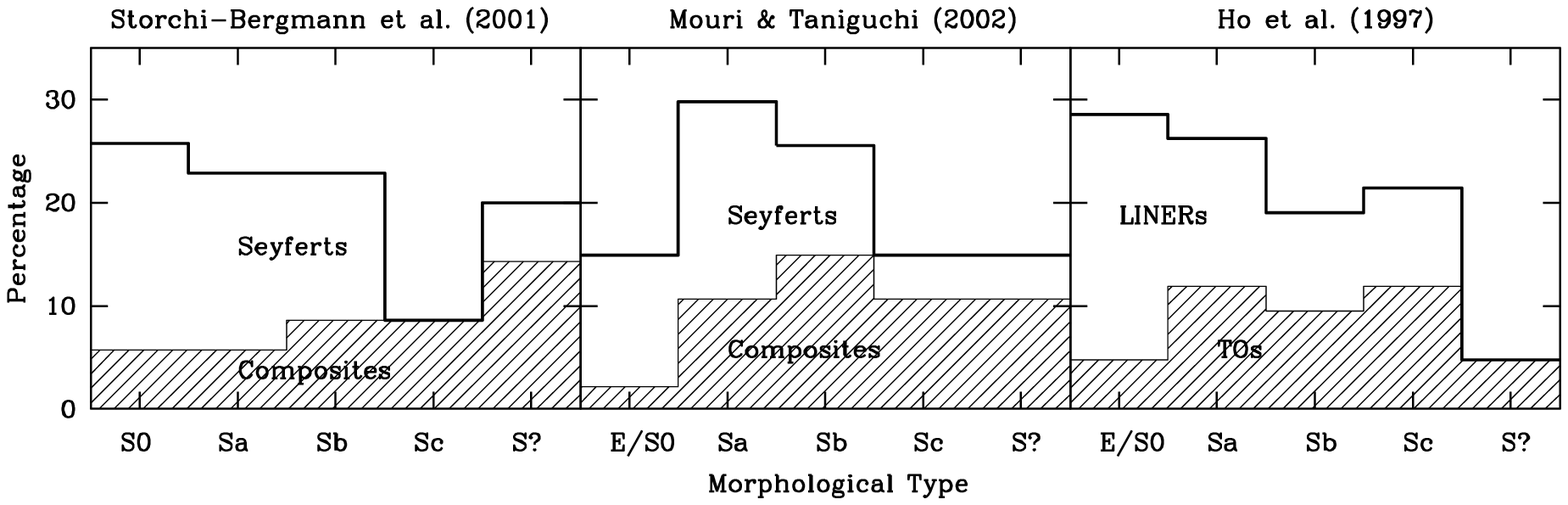}}
\caption{Distribution of Hubble types of:
Seyfert 2s from \cite{SB01} (left), CfA Seyferts from \cite{MT02}
(middle), and LINERs from \cite{Cid04}  and \cite{R04} (right). 
Open histograms represent the distributions of the
complete samples, while hatched ones represent the distribution
of galaxies with large contribution from young (first two
samples) and intermediate age stars (last sample).}\label{fig}
\end{figure}

\section{Conclusions}

In figure~\ref{fig} we compare the distribution of Hubble types
for the Seyfert and LINER samples discussed above, and the corresponding
subsamples of Composites and TO's. This figure shows that
both Composites and TO's present a larger incidence in later-type
galaxies, although the distributions are not
exactly the same, with the Seyfert samples having a larger
proportion of distorted galaxies (S?).   

We quantified the contribution from young and intermediate age stars to
the spectra, and find that the contribution in luminosity of the young stars
in the Composites and of the intermediate age stars to the
TO's correspond to similar mass contributions
to the total mass of their bulges.

We thus conclude that at least part of the TO's may be evolved Composites;
in this case, the Seyfert 2 nucleus should have evolved and faded to a LINER 
together with the Starburst. This conclusion is also consistent
whith non-Composite Seyfert nuclei evolving to LINER nuclei (Corbin, 2000).
Our results also support that interactions may be the triggers of
starburst activity, once the interaction signatures
are gone after 10$^9$yrs.


\begin{thebibliography}{}

\bibitem[Cid Fernandes et al. 2001]{Cid01}
     {Cid Fernandes, R., et al.} 2001, ApJ, 558, 81

  \bibitem[Cid Fernandes et al. 2004]{Cid04}
     {Cid Fernandes, R., et al.} 2004, ApJ, 605, 105

 \bibitem[Corbin(2000)]{2000ApJ...536L..73C}
 Corbin, M.~R., 2000, ApJ, 536, L73

 \bibitem[Gonz\'alez Delgado et al. 2004]{R04}
     {Gonz\'alez Delgado, R., et al.} 2004, ApJ, 605, 127

 \bibitem[Ho et al. 1997]{Ho97}
     {Ho, L. C., Filippenko, A. V., \& Sargent, W. L. W.} 1997, ApJ, 487, 568

  \bibitem[Malkan et al. 1998]{M98}
     {Malkan, M., et al.} 1998, ApJS, 117, 25

  \bibitem[Mouri \& Taniguchi (2002)]{MT02}
     {Mouri, H., \& Taniguchi, Y.} 2002, ApJ, 565, 786

  \bibitem[Storchi-Bergmann et al. (2001)]{SB01}
     {Storchi-Bergmann, T., et al.} 2001, ApJ, 559, 147

\bibitem[Schmitt(2001)]{schmitt01} Schmitt, H. R. 2001, AJ, 122, 2243

\end{thebibliography}
\end{document}